\pdfoutput=1

\documentclass[11pt,table,dvipsnames]{article}

\usepackage[preprint]{acl}
\usepackage{amsmath}
\usepackage{times}
\usepackage{latexsym}
\usepackage{xspace}

\usepackage[T1]{fontenc}

\usepackage[utf8]{inputenc}

\usepackage{microtype}

\usepackage{inconsolata}

\usepackage{graphicx}

\usepackage{graphicx}
\usepackage{subfigure}
\usepackage{caption}
\usepackage{hyperref}
\usepackage{xcolor}
\usepackage{multirow}
\usepackage{float}
\usepackage{booktabs}
\usepackage{algorithmic}
\usepackage[algo2e,ruled,linesnumbered]{algorithm2e}
\definecolor{light-gray}{gray}{0.95}
\newcommand{\tool}{\textsc{Solla}\xspace}

\title{Instructions for *ACL Proceedings}
\title{\tool: Towards a Speech-Oriented LLM That Can Listen While Talking}
\title{\tool: Towards a Speech-Oriented LLM That Hears Acoustic Context}

\author{Junyi Ao$^1$, Dekun Chen$^1$, Xiaohai Tian$^2$, Wenjie Feng$^1$,\\
\textbf{Jun Zhang}$^2$, \textbf{Lu Lu}$^2$, \textbf{Yuxuan Wang}$^2$, \textbf{Haizhou Li}$^1$, \textbf{Zhizheng Wu}$^1$\thanks{Corresponding author: wuzhizheng@cuhk.edu.cn}  \\
$^1$ School of Data Science, Shenzhen Research Institute of Big Data, \\
The Chinese University of Hong Kong, Shenzhen \\
$^2$Bytedance
}

\begin{document}
\maketitle
\begin{abstract}
Large Language Models (LLMs) have recently shown remarkable ability to process not only text but also multimodal inputs such as speech and audio.
However, most existing models primarily focus on analyzing input signals using text instructions, \textit{overlooking scenarios in which speech instructions and audio are mixed and serve as inputs to the model}.
To address these challenges, we introduce \tool, a novel framework designed to understand speech-based questions and hear the acoustic context concurrently.
\tool incorporates an audio tagging module to effectively identify and represent audio events, as well as an ASR-assisted prediction method to improve comprehension of spoken content.
To rigorously evaluate \tool and other publicly available models, we propose a new benchmark dataset called SA-Eval, which includes three tasks: audio event classification, audio captioning, and audio question answering.
SA-Eval has diverse speech instruction with various speaking styles, encompassing two difficulty levels, \textit{easy} and \textit{hard}, to capture the range of real-world acoustic conditions.
Experimental results show that \tool performs on par with or outperforms baseline models on both the \textit{easy} and \textit{hard} test sets, underscoring its effectiveness in jointly understanding speech and audio.\footnote{SA-Eval is available at \url{https://github.com/amphionspace/SA-Eval}}

\end{abstract}

\section{Introduction}
People encounter a wide variety of sounds in their daily lives, such as speech, ambient noises, and music.
Each carries important information that helps us interact with our environment.
For instance, when you and your friend are in a forest, you might ask, ``Do you recognize this bird sound?''
Such simple exchanges are part of everyday life.
As we navigate these everyday experiences, our ability to interpret and respond to diverse sound cues is fundamental to effective communication and situational awareness.
We expect a speech-oriented LLM to do the same thing as our friend.
In this context, developing speech-oriented LLMs that can seamlessly integrate and understand these varied auditory inputs is a technical challenge and a critical step toward enhancing human-computer interaction.

\begin{figure}
    \centering
    \includegraphics[width=0.48\textwidth]{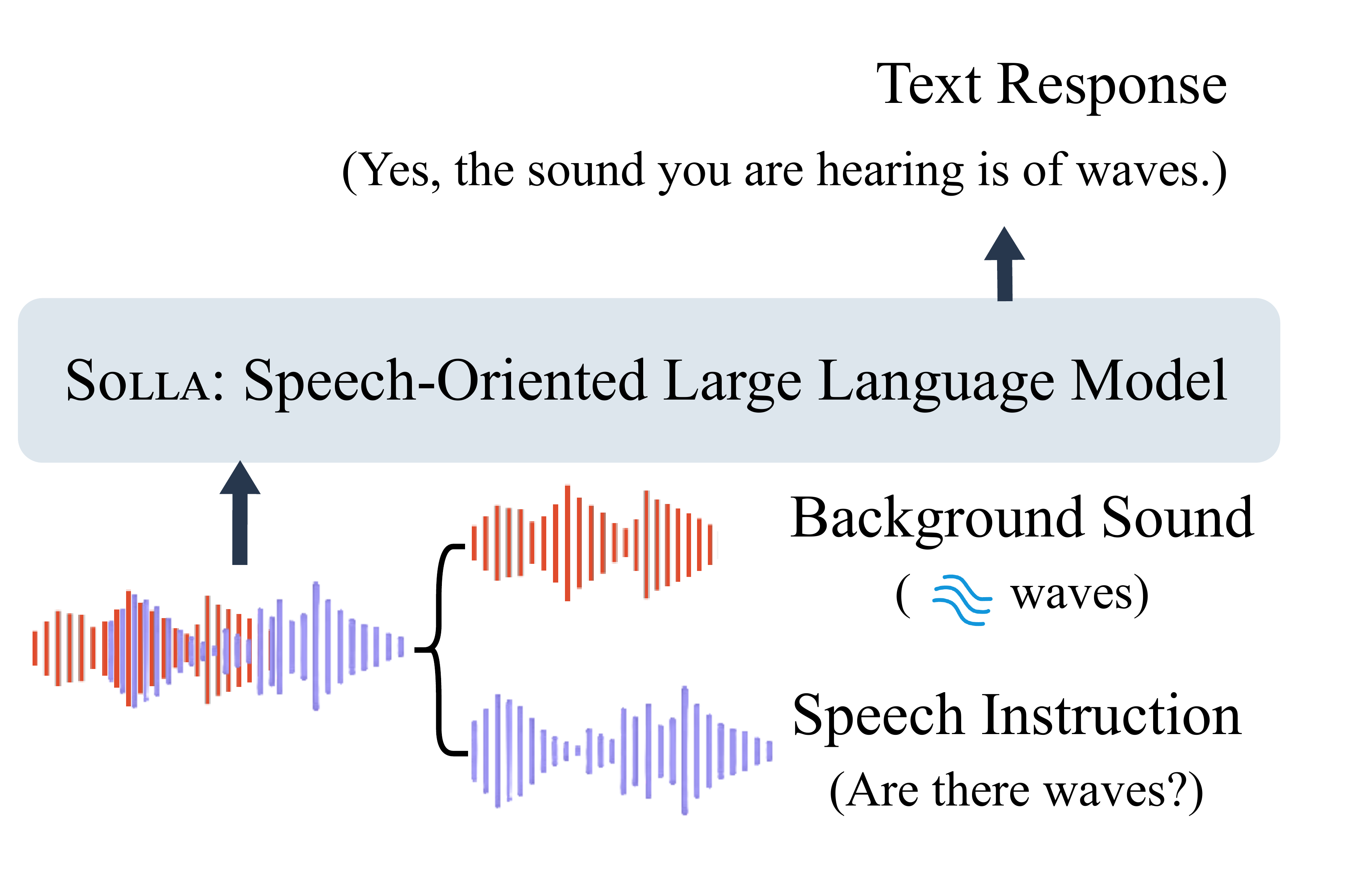}
    \caption{An illustration of \tool that a speech-oriented LLM can understand the acoustic context and generate the answer.\label{fig:task}}
\end{figure}

LLMs have demonstrated extraordinary versatility as a universal interface for a broad range of assistance \cite{achiam2023gpt,team2023gemini,touvron2023llama1,touvron2023llama2,yang2024qwen2,team2024gemma}.
Recently, their capabilities have expanded beyond text, allowing them to interpret multimodal inputs, such as speech \cite{chu2023qwen,chu2024qwen2,tang2024salmonn,hu2024wavllm,openai2024gpt4o,defossez2024moshi,fang2024llama,xie2024mini}, which broadens the range of tasks LLMs can address.
These speech LLMs have achieved impressive results on various tasks, such as speech recognition and speech translation.
They primarily focus on understanding input signals and performing tasks guided by text instructions \cite{chu2023qwen,chu2024qwen2,tang2024salmonn,hu2024wavllm}.
Recently, many studies \cite{defossez2024moshi,fang2024llama,xie2024mini} have begun exploring seamless voice interaction to enable more natural and low-latency human-computer communication.
However, despite these efforts, few of these models currently address the simultaneous understanding of audio and speech.
Some previous works \cite{kong2024audio,gong2024listen,deshmukh2023pengi} also focus on analyzing and processing audio input.
They usually use an audio encoder to extract audio features that serve as input for the LLM.
These models still require text instructions to define the corresponding tasks. \textit{In a hand-free speech interaction, it is infeasible to provide text instructions}.

It is a challenging task to understand speech instructions with acoustic background context (e.g. audio events, background speech, music).
First, the model should be able to accurately identify the question from a speech instruction.
Second, an appropriate response needs to be formulated based on the information conveyed by the acoustic context.
Moreover, real-world scenarios often involve complex mixtures of speech and background sound.
For instance, overlapping audio sources and varying signal-to-noise ratios (SNR) can make it difficult for the model to extract the essential information in both the speech and surrounding audio.
As a result, the model may need to disentangle these different sources before providing a reliable answer.

As an initial step towards understanding speech instructions and background acoustic context, this work focuses on audio events context.
We \textbf{propose a \tool model, a speech-oriented LLM that hears acoustic context, and design a SA-Eval benchmark dataset}.
To help the model better interpret audio information from input signals, we introduce an audio tagging module that predicts audio events and generates corresponding event representations.
We also present an ASR-assisted prediction task, which allows the model to first capture the spoken content before producing its response.

To evaluate the performance of \tool and publicly available models, we introduce SA-Eval, which is built on publicly available datasets and consists of three tasks: audio event classification, audio captioning, and audio question answering. According to the varied challenges of the speech instructions, we divide SA-Eval into two different difficulty levels, \textit{easy} and \textit{hard}, to facilitate a more comprehensive evaluation.
In the \textit{easy} mode test set, the speech instructions and audio do not overlap, whereas in the \textit{hard} mode test set, we simulate a real-world scenario where the speech instructions and audio overlap with low SNR value, in order to assess whether the model can accurately capture the speech instruction information in noisy environments.

\section{\tool Framework}

\tool deals with speech instructions and audio context understanding. 
We formulate the task as follows:
given a speech mixture
\begin{equation}
    s_m = s_s + s_a
\end{equation}
where $s_s$ represents the speech instruction and $s_a$ denotes the accompanying audio signal.
The model must generate a text response that integrates both types of information.
For example, the speech instruction may prompt the model to predict an event occurring in the audio or ask a related question. The subsequent sections detail each module.

\subsection{Overview}

\begin{figure}
    \centering
    \includegraphics[width=0.45\textwidth]{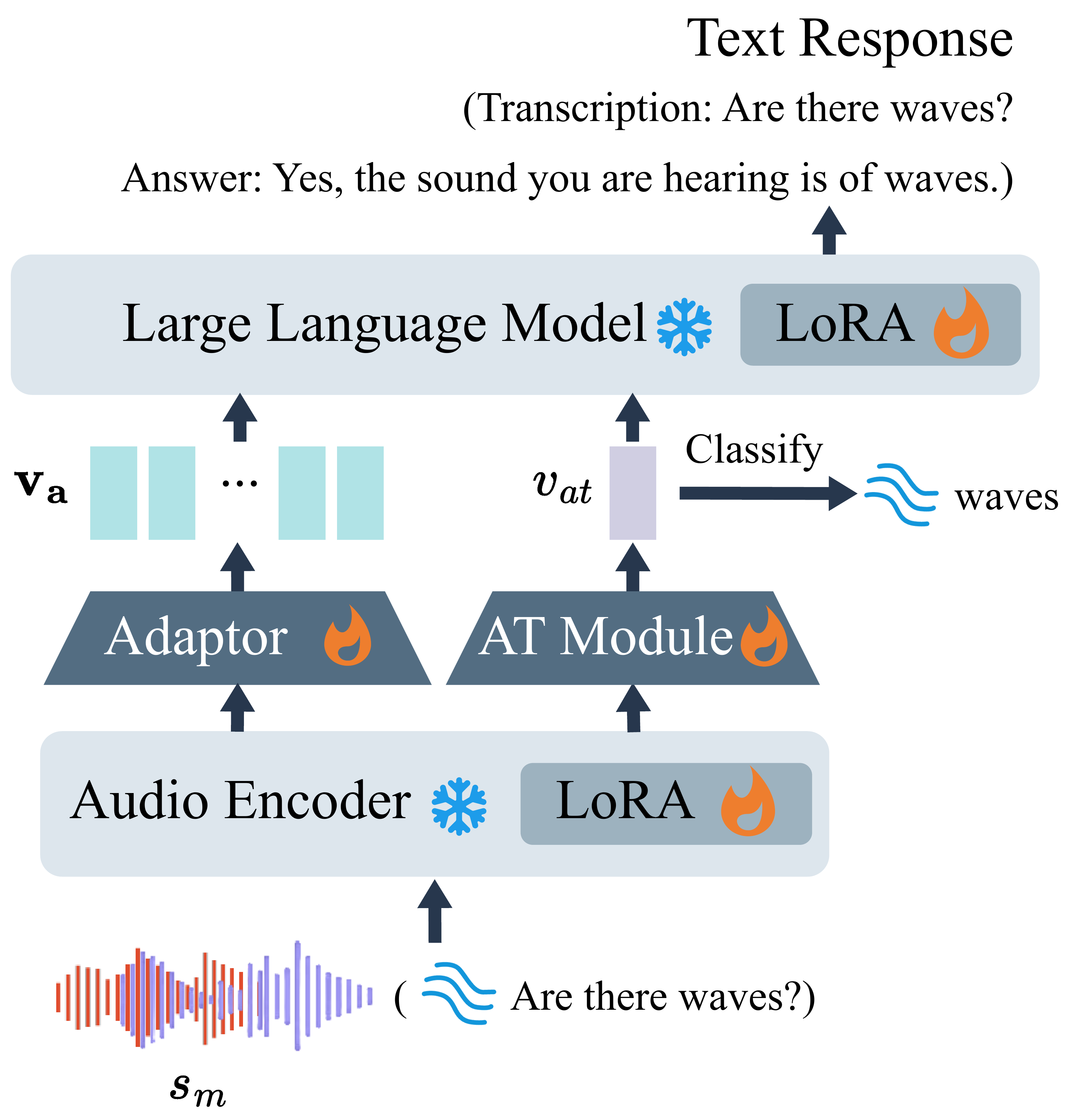}
    \caption{The overview framework of \tool. AT module denotes the audio tagging module. \label{fig:model}}
\end{figure}

\tool is a speech-oriented large language model.
The framework of \tool is illustrated in Figure \ref{fig:model}. The \tool is built upon four key components: an audio encoder, an adaptor, an Audio Tagging (AT) module, and a large language model (LLM).
In addition to generating the corresponding answer, \tool also has an additional ASR-assisted prediction task to aid in understanding speech instructions.

\subsection{Audio Encoder}

The model's audio encoder is the encoder part of Whisper-Large-v3 \cite{radford2022whisper}.
We choose the Whisper encoder as the audio encoder for two reasons.
First, whisper is trained on 680,000 hours of multilingual data with multitask supervision, which yields high accuracy and robustness in speech recognition.
Consequently, its encoder is widely adopted for processing speech data \cite{chu2023qwen,chu2024qwen2,tang2024salmonn}.
Second, recent work \cite{gong2023whisper} also demonstrates that its representations are noise-variant, capturing rich audio information.
Accordingly, we use the Whisper encoder to efficiently extract speech and audio information from the input speech mixture $s_m$.
During training, we freeze the encoder parameters and fine-tune the audio encoder using a trainable LoRA adaptor \cite{hu2021lora}.

\subsection{Adaptor}

The adaptor is designed to transform the raw output from the audio encoder into a suitable format that the LLM can process effectively.
Specifically, it comprises two linear layers: the first is followed by a GELU activation function \cite{hendrycks2016gaussian}, and the second is succeeded by a two-dimensional average pooling operation that downsamples the representation by a factor of two.
The final output is a sequence representation $\mathbf{v_a} = (v_a^1, v_a^2, ..., v_a^n)$, where $n$ denotes its length.

\subsection{AT Module}

To capture audio-specific information, we introduce a novel Audio Tagging (AT) module.
This module begins by applying a two-dimensional average pooling operation to the audio encoder outputs, reducing the temporal resolution by a factor of twenty.
The pooled features are then normalized via layer normalization and transformed through a linear projection to yield an intermediate representation.
A residual attention block subsequently models long-range temporal dependencies.
After mean-pooling along the temporal dimension, the resulting compact feature vector $v_{at}$ is normalized and passed through a linear classifier to produce logits for audio event prediction.

\subsection{Large Language Model}
We employ the InternLM2-chat model (InternLM2-chat-7b) \cite{cai2024internlm2}, a 7-billion-parameter large language model optimized for dialogue generation.
During training, the LLM remains frozen while a trainable LoRA adaptor \cite{hu2021lora} is employed to enable fine-tuning.
The LLM takes the adaptor's output representation $(v_a^1, v_a^2, ..., v_a^n)$ and the AT module's pooled vector $v_{at}$ as its inputs.

\subsection{ASR-Assisted Prediction}
To enhance the model's comprehension of speech instructions, we incorporate an ASR-assisted prediction strategy.
During training, the model first performs the ASR task before generating the final answer.
For example, the target output may be \textit{Transcription: Are there waves? Answer: Yes, the sound you are hearing is of waves}, as shown in Figure \ref{fig:model}.
This approach ensures that the model accurately captures the content of the speech instruction, which is then combined with the audio information to produce the final response.

\section{SA-Eval Benchmark Dataset}
\label{sec:sa_eval}
\subsection{Overview}

The SA-Eval dataset is derived from test sets of widely used datasets spanning three different tasks, as shown in Table \ref{tab:dataset}.
To ensure diversity in both question formulation and vocal characteristics of speech instructions, we leverage GPT-4o \cite{openai2024gpt4o} to generate multiple question formats for each task while employing TTS models to synthesize varied speaking styles.
Below, we provide a detailed explanation of the dataset construction process.

\begin{table*}
\centering
\caption{Statistics of the SA-Eval benchmark dataset, which is built upon a few open-source datasets for audio classification, audio captioning and audio QA. We reuse them and mix them with speech instructions for SA-Eval. In addition, we divide the SA-Eval into \textit{easy} and \textit{hard} subsets based on SNR. \label{tab:dataset}}
\resizebox{\textwidth}{!}{
\begin{tabular}{lcccccc}
    \toprule
    \multirow{2}{*}{\textbf{Dataset}} &  \multirow{2}{*}{\textbf{Task}}  & \multirow{2}{*}{\textbf{Average Duration (s)}}  &  \multirow{2}{*}{\textbf{\# QA Pairs}}  & \multirow{2}{*}{\textbf{\# Instructions}}  & \multicolumn{2}{c}{\textbf{SNR}} \\
     \cmidrule(l){6-7}
    & & & & & \textbf{easy} & \textbf{hard} \\
    \midrule
    \midrule
    VGGSound \cite{chen2020vggsound} & Audio Classification & 12.4 & 15,341 & 402 & -0.8 & -10.0 \\
    AudioSet \cite{gemmeke2017audio} & Audio Classification & 12.3 & 17,241 & 402 & -2.3 & -11.8 \\
    FSD50K \cite{fonseca2021fsd50k} & Audio Classification & 12.6 & 10,231 & 402 & -0.9 & -12.6 \\
    AudioCaps \cite{kim2019audiocaps} & Audio Captioning & 12.2 & 881 & 345 & -1.8 & -11.5  \\
    Clotho V2 \cite{9052990} & Audio Captioning & 24.3 & 1,045 & 355   & -6.1 & -14.4 \\
    Clotho-AQA \cite{lipping2022clotho} & Audio Question Answering & 23.6 & 8,430 & 2515 & -9.7 & -17.4 \\
     \bottomrule
\end{tabular}
}
\end{table*}

\subsection{Base Datasets}
\label{sec:data_col}
We construct the SA-Eval dataset based on the test sets of six widely used datasets, including VGGSound \cite{chen2020vggsound}, AudioSet \cite{gem2017audioset}, FSD50K \cite{fonseca2021fsd50k}, AudioCaps \cite{kim2019audiocaps}, Clotho V2 \cite{9052990} and Clotho-AQA \cite{lipping2022clotho}, encompassing three different tasks, audio classification, audio captioning and audio question answering. 
Please note that several test sets, including VGGSound, AudioSet, and AudioCaps, are missing some audio files due to data corruption or unavailability of the source files.

\subsection{Generating Text Instructions}
\label{Section 3.3}
Except for Clotho-AQA, which includes questions corresponding to each audio, we generate diverse instructions for the other two tasks.
Specifically, we begin by formulating multiple instructions for these tasks and then use a rewriting prompt with GPT-4o \cite{openai2024gpt4o} to produce 100 unique sentences for each instruction.
Upon reviewing the outputs, we find that some generated instructions are ill-suited for the target tasks.
As a result, we engage human annotators to filter out any unsuitable instructions.
Details can be found in Appendix \ref{apd:ins_gen}.
Table \ref{tab:dataset} shows how many unique instructions each test set contains finally.

\subsection{Generating Speech Instructions}

We generate speech for our instructions by leveraging MaskGCT \cite{wang2024maskgct} together with Microsoft Azure TTS \footnote{\url{https://learn.microsoft.com/en-us/azure/ai-services/speech-service/text-to-speech}}.
Specifically, we select speakers from the Emilia dataset \cite{he2024emilia} who have at least seven speech recordings.
We then retain the top seven recordings ranked by DNSMOS scores \cite{reddy2021dnsmos}.
This process yields a pool of 9,880 speakers and 69,160 audio recordings that are randomly used as speech prompts for MaskGCT.

Following speech generation with MaskGCT, we observe that a small fraction of the speech exhibits issues, such as excessive length, content inaccuracies, or language errors.
To address this, we further filter these problematic recordings based on the content and regenerate this subset of speech using Microsoft Azure TTS.
Specifically, we first calculate the word error rate (WER) for each utterance generated by MaskGCT.
Those with a WER exceeding 40\% are regenerated, while human annotators manually review utterances with lower WERs to determine if regeneration is necessary.s
The speaker types for Microsoft Azure TTS are randomly selected.

\subsection{Mixing Audio Events and Speech Instructions}
In real-world scenarios, speech instructions and accompanying audio can appear in various configurations.
For instance, the speech instruction might precede the audio, or the two might overlap.
To simulate these diverse conditions and evaluate model performance across different difficulty levels, we generate test sets in two modes, \textit{easy} and \textit{hard}, as detailed in Algorithm \ref{alg:1} of Appendix \ref{apd:mix_algo}.
In our algorithm, the \textit{easy} mode features non-overlapping audio with a higher SNR, whereas the \textit{hard} mode involves overlapping audio with a lower SNR, thereby mimicking challenging real-world situations where high audio volume can obscure speech instructions.

To create these datasets, we first decide whether to overlap the audio signal $s_a$ and the speech instruction $s_s$, applying appropriate padding when necessary.
In the \textit{hard} mode, $s_a$ and $s_s$ always overlap, while in the \textit{easy} mode, they remain non-overlapping.
Second, we sample the loudness of both $s_a$ and $s_s$ using loudness units relative to full scale (LUFS) \cite{series2011algorithms}, following the mixing strategy of LibriMix \cite{cosentino2020librimix}.
LUFS, defined by the ITU-R BS.1770-4 recommendation \cite{series2011algorithms}, better reflects perceived loudness than traditional SNR metrics, as it is unaffected by silence and remains relatively insensitive to downsampling \cite{cosentino2020librimix}.
For the \textit{easy} mode, both signals are uniformly sampled between -30 and -25 LUFS.
In the \textit{hard} mode, $s_a$ is sampled between -23 and -20 LUFS, while $s_s$ is sampled between -33 and -30 LUFS to simulate scenarios where the speech instruction is less clear.
When necessary, the signals are clipped to 0.9 before being mixed to produce the final mixture $s_m$.

\section{Experimental Setups}
\subsection{Training Set Preparation}
\label{sec:training}
We prepare the training set using a procedure similar to that of the SA-Eval dataset.
Specifically, we construct the set using the training data from FSD50K \cite{fonseca2021fsd50k}, AudioSet\cite{gem2017audioset}, AudioCaps \cite{kim2019audiocaps}, FreeSound\cite{10.1145/2502081.2502245}, Sound Bible \cite{soundbible}, VGGSound \cite{chen2020vggsound}, Clotho V2 \cite{9052990}, Clotho-AQA \cite{lipping2022clotho}, and MusicQA \cite{liu2024music}.
For all datasets except Clotho-AQA and MusicQA, we use the instructions from OpenAQA \cite{gong2024listen}.
There are several differences compared to the preparation of the SA-Eval dataset.

First, we generate the corresponding speech instructions using MaskGCT \cite{wang2024maskgct} without additional filtering.
Second, the parameters for speech instruction and audio mixing are different:
80\% of the speech instructions and audio overlapped, and we randomly set the loudness of speech instructions between -38 and -20 LUFS, while adjusting the audio loudness between -25 and -33 LUFS.
Finally, we replace 20\% of the speech instructions with those from SA-Eval for audio captioning and audio classification tasks.
Detailed data statistics of the training set are provided in Appendix \ref{apd:trainin_set}.

\subsection{Configurations for Model Training}
All implemented models are developed using Xtuner \cite{2023xtuner}.
The optimization is performed with the AdamW optimizer \cite{loshchilov2017decoupled}, employing a learning rate of $2 \times 10^{-4}$.
Model fine-tuning is conducted over two epochs on 16 A100 GPUs, each processing a batch size of 16, with each experiment taking approximately four days.
For the LoRA adapter of the InternLM2-chat-7b model, we set the rank to 512 and $\alpha$ to 256. In contrast, for the encoder of the Whisper large-v3 model, the rank is set to 64 and $\alpha$ to 16.

\subsection{Baseline Models}
In addition to \tool, we implement two baseline models for comparison, namely Audio LLM$_{\text{text}}$ and Audio LLM$_{\text{speech}}$.
Unlike \tool, these two models consist only of an audio encoder, an adaptor, and an LLM.
They use the answer as the target without ASR-assisted prediction.
All other training parameters remain the same as \tool.

The first baseline, Audio LLM$_{\text{text}}$, takes audio and text instruction as input.
This model allows us to assess the effect of using a speech versus a text instruction prompt on performance.
In contrast, Audio LLM$_{\text{speech}}$ receives the same input as \tool. %

\subsection{Publicly Available Models}

We compare two types of models: one guided by text instructions and the other by speech instructions.
For models using text instruction, we strictly follow the original studies' evaluation protocols to ensure fairness, focusing only on the in-distribution test sets that correspond to their training data.
For models that do not report their training sets, including Qwen-Audio \cite{chu2023qwen} and Qwen2-Audio \cite{chu2024qwen2}, we selectively compare the test sets or tasks evaluated in their papers.
In contrast, for models using speech instruction, we employ SA-Eval for evaluation, as detailed in Section \ref{sec:sa_eval}.
Overall, our analysis includes the following publicly available models for comparison:

\begin{table*}
\centering
\caption{Main results of publicly available models and self-implemented models across six test sets of three tasks.
For all metrics, larger numbers indicate better performance.
The dash indicates that the model is not tested on the relevant dataset in the original paper.
\dag Pengi's AudioCaps results come from the original paper. Due to data corruption or unavailability of the source files, the test data used may be different from that used by other models, as explained in Section \ref{sec:data_col}. \label{tab:main_res}}
\resizebox{\textwidth}{!}{
\begin{tabular}{lccccccccccc}
    \toprule
    \multirow{3}{*}{\textbf{Model}} & \multicolumn{3}{c}{\textbf{Audio Classification}} & \multicolumn{2}{c}{\textbf{Audio Captioning}} & \multicolumn{1}{c}{\textbf{Audio Question Answering}} \\
    \cmidrule(l){2-4} \cmidrule(l){5-6}  \cmidrule(l){7-7}
    & \textbf{VGGSound} & \textbf{AudioSet} & \textbf{FSD50K} & \textbf{Clotho V2} & \textbf{AudioCaps} & \textbf{Clotho-AQA} \\
    &ACC  & F1  & F1  & CIDEr / SPICE  / SPIDEr  & CIDEr / SPICE  / SPIDEr  & ACC$_{\text{All}}$ / ACC$_{\text{Una}}$ \\
    \midrule
    \midrule
    \rowcolor{light-gray}
    \multicolumn{7}{c}{\textbf{\textit{Models using Text Instruction}}} \\
    \midrule
    Pengi \dag \cite{deshmukh2023pengi} & - & - & 52.6 & 41.6 / 12.6 / 27.1 & 75.2 / 18.2 / 46.7 & 46.4 / 62.3 \\
    Audio Flamingo \cite{kong2024audio} & - & - & 56.0 & 46.5 / 12.8 / 29.8 & - / - / - & 58.5 / 86.9  \\
    LTU \cite{gong2024listen} & 53.6 & 28.3 & 47.8 & 31.6 / 12.0 / 21.8 & 48.5 / 16.8 / 32.5 & - / - \\
    Qwen-Audio \cite{chu2023qwen} & - & - & - & 44.1 / 13.6 / 28.8 & - / - / - & 57.9 / 74.9 \\
    Qwen2-Audio-AA \cite{chu2024qwen2} & - & - & - & 32.8 / 10.8 / 21.8 & 47.9 / 16.5 / 32.2 & - / -\\
    Audio LLM$_\text{text}$ & 55.6 & 26.0 & 57.4 & 39.4 / 12.5 / 25.9 & 62.7 / 15.6 / 39.2 & 62.2 / 87.9  \\
    \midrule
    \rowcolor{light-gray}
    \multicolumn{7}{c}{\textbf{\textit{SA-Eval - easy mode}}} \\
    \midrule
    Qwen2-Audio-VC \cite{chu2024qwen2} & 10.9 & - & - & - / - / - & - / - / - & 45.7 / 59.2 \\
    GPT-4o \cite{openai2024gpt4o} & 7.7  & - & - & - / - / - & - / - / - & 45.8 / 56.9 \\
    Audio LLM$_\text{speech}$ & 53.4 & 25.2 & 53.1 & 38.7 / 12.4 / 25.6 & 64.0 / 15.5 / 39.8 & 61.7 / 86.8 \\
    \tool & 53.8 & 25.2 & 55.1 & 38.5 / 12.6 / 25.6 & 63.2 / 16.0 / 39.6 & 62.1 / 86.9 \\
    \midrule
    \rowcolor{light-gray}
    \multicolumn{7}{c}{\textbf{\textit{SA-Eval - hard mode}}} \\
    \midrule
    Qwen2-Audio-VC \cite{chu2024qwen2} & 18.3 & - & - & - / - / - & - / - / - & 34.3 / 46.2 \\
    GPT-4o \cite{openai2024gpt4o} & 4.8 & - & - & - / - / - & - / - / - & 35.2 / 39.1 \\
    Audio LLM$_\text{speech}$ & 53.7 & 23.7 & 51.6 &  37.2 / 12.1 / 24.6 & 58.8 / 14.5 / 36.7 & 54.8 / 75.7 \\
    \tool & 53.5 & 23.8 & 54.0 & 38.6 / 12.3 / 25.5 & 60.6 / 15.4 / 38.0 & 55.4 / 78.3 \\
     \bottomrule
\end{tabular}
}
\end{table*}

\paragraph{Pengi \cite{deshmukh2023pengi}} Pengi reformulates both open- and closed-ended audio tasks as text-generation problems. It uses transfer learning and instruction-tuned templates to process audio and text without task-specific fine-tuning.

\paragraph{Audio Flamingo \cite{kong2024audio}} Audio Flamingo extends large language models to interpret diverse audio, including non-speech sounds. It excels in audio comprehension, few-shot learning, and multi-turn dialogues. %

\paragraph{LTU \cite{gong2024listen}} LTU fuses audio perception with reasoning. Trained on the OpenAQA-5M dataset using a perception-to-understanding curriculum, it outperforms rivals in audio classification and captioning tasks.

\paragraph{Qwen-Audio \cite{chu2024qwen2}} Qwen-Audio is pre-trained on over 30 tasks, including speech, ambient sounds, and music, in multiple languages.
Its broad training yields strong performance without specialized fine-tuning.

\paragraph{Qwen2-Audio \cite{chu2024qwen2}} Building on Qwen-Audio, Qwen2-Audio is trained on larger datasets and optimized with Direct Preference Optimization \cite{rafailov2024direct} for better human alignment.
It offers two modes: Voice Chat (Qwen2-Audio-VC) for dialogue using speech instruction and Audio Analysis (Qwen2-Audio-AA) for detailed processing using text instruction.

\paragraph{GPT-4o \cite{openai2024gpt4o}} GPT-4o \footnote{The version of GPT-4o is \texttt{gpt-4o-audio-preview-2024-12-17}.} combines advanced language understanding with enhanced speech recognition and generation, supporting tasks from creative writing to real-time conversation.

\subsection{Evaluation Metrics and Methods}
For single-label audio classification and audio question answering, we gauge performance with accuracy (\%).
We report two results on the Clotho-AQA test set.
The first, ACC$_{\text{All}}$, represents the overall accuracy across all test data.
The second, ACC$_{\text{Una}}$, measures the accuracy on questions where the answers from all three annotators are consistent, with "Una" standing for unanimous.
For multi-label audio classification, we adopt the F1 score, applying the macro averaging approach.
For audio captioning, we rely on CIDEr \cite{vedantam2015cider}, SPICE \cite{10.1007/978-3-319-46454-1_24}, and SPIDEr \cite{Liu_2017_ICCV}. %

For single-label audio classification and audio question answering tasks, if the model's recognized answer exactly matches the ground truth, we consider it correct. 
Otherwise, we rely on o1-mini\footnote{The version of o1-mini is \texttt{o1-mini-2024-09-12}.} to assess whether the output is correct.
The specific prompts can be found in the Appendix \ref{apd:eval_jud}.
For multi-label audio classification, if the model's predicted label does not match any of the known labels, we calculate the cosine similarity between the text embedding \footnote{The model for extracting text embedding is \texttt{text-embedding-3-large}.} of the predicted label and those of all the known labels, and select the label with the highest similarity as the predicted label.
\section{Experimental Results and Analysis}
\begin{table*}
\centering
\caption{Ablation Study for AT Module and ASR-assisted prediction on SA-Eval. ``w/o'' refers to without. For all metrics, larger numbers indicate better performance. \label{tab:ablation}}
\resizebox{\textwidth}{!}{
\begin{tabular}{lccccccc}
    \toprule
    \multirow{3}{*}{\textbf{Model}} & \multicolumn{3}{c}{\textbf{Audio Classification}} & \multicolumn{2}{c}{\textbf{Audio Captioning}} & \multicolumn{1}{c}{\textbf{Audio Question Answering}} \\
    \cmidrule(l){2-4} \cmidrule(l){5-6}  \cmidrule(l){7-7}
    & \textbf{VGGSound} & \textbf{AudioSet} & \textbf{FSD50K} & \textbf{Clotho V2} & \textbf{AudioCaps} & \textbf{Clotho-AQA} \\
    &ACC  & F1  & F1  & CIDEr / SPICE  / SPIDEr  & CIDEr / SPICE  / SPIDEr  & ACC$_{\text{All}}$ / ACC$_{\text{Una}}$ \\
    \midrule
    \midrule
    \rowcolor{light-gray}
    \multicolumn{7}{c}{\textbf{\textit{SA-Eval - easy mode}}} \\
    \midrule
    \tool & 53.8 & 25.2 & 55.1 & 38.5 / 12.6 / 25.6&63.2 / 16.0 / 39.6&62.1 / 86.9  \\
    \quad w/o AT Module & 53.1&24.4&54.1&38.4 / 12.8 / 25.6&64.2 / 16.3 / 40.3&61.5 / 85.8 \\
    \quad w/o ASR-Assisted Pred & 54.1&24.4&53.3&40.0 / 12.5 / 26.3&62.8 / 15.5 / 39.2&61.4 / 86.3 \\
    \quad w/o Both (Audio LLM$_\text{speech}$) & 53.4 &25.2&53.1&38.7 / 12.4 / 25.6&64.0 / 15.5 / 39.8&61.7 / 86.8 \\
    \midrule
    \rowcolor{light-gray}
    \multicolumn{7}{c}{\textbf{\textit{SA-Eval - hard mode}}} \\
    \midrule
    \tool & 53.5& 23.8& 54.0& 38.6 / 12.3 / 25.5& 60.6 / 15.4 / 38.0& 55.4 / 78.3 \\
    \quad w/o AT Module &53.0&23.1&52.9&38.1 / 12.4 / 25.2&61.6 / 15.7 / 38.7&54.8 / 76.4 \\
    \quad w/o ASR-Assisted Pred & 54.1&23.3&51.8&37.5 / 12.1 / 24.8&58.7 / 14.7 / 36.7&55.6 / 76.9 \\
    \quad w/o Both (Audio LLM$_\text{speech}$) & 53.7&23.7&51.6&37.2 / 12.1 / 24.6&58.8 / 14.5 / 36.7&54.8 / 75.7 \\
     \bottomrule
\end{tabular}
}
\end{table*}
\subsection{Main Results}
Table \ref{tab:main_res} shows the results of both publicly available models and our self-implemented models across six test sets of three tasks.
In all cases, higher metric values correspond to better performance.

First, the experimental results show that \tool performs on par with or exceeds Audio LLM$_{\text{speech}}$ on SA-Eval. %
In particular, \tool exhibits significantly better results than Audio LLM$_{\text{speech}}$ in \textit{hard} mode across most test sets.
This highlights the robustness of \tool across various scenarios.

Second, Audio LLM$_{\text{text}}$ provides a solid baseline for comparison as it outperforms other publicly available models on VGGSound, FSD-50K, and Clotho-AQA.
Its performance drops on other test sets, possibly due to the training data.
As detailed in Section \ref{sec:training}, the majority of its training instructions come from OpenAQA \cite{gong2024listen}, which mainly offers open-ended instructions.
Consequently, Audio LLM$_{\text{text}}$ also outperforms LTU which is trained using OpenAQA across nearly all test sets.

Third, comparing the results of \tool and Audio LLM$_{\text{text}}$ reveals that \tool's performance has declined across all test sets, with a more pronounced drop in \textit{hard} mode.
This indicates that even with the assistance of the AT Module and ASR-assisted prediction, the model still has room for improvement in understanding speech instructions and audio information.

Finally, we compare \tool with two models using speech instruction, Qwen2-Audio-VC and GPT-4o.
We focus our evaluation on VGGSound and Clotho-AQA for two reasons.
First, the metrics used for audio captioning do not account for synonyms \cite{gong2024listen}, so comparing results for models not trained on the relevant datasets would be unfair.
Second, as single-label audio classification and audio question answering datasets, VGGSound and Clotho-AQA better reflect a model's ability to recognize audio events and comprehend audio information, unlike multi-label audio classification requiring multiple audio event outputs and more complicated evaluation.
Experimental results show that \tool significantly outperforms GPT-4o and Qwen2-Audio-VC in both evaluation modes.

\subsection{Ablation Study}
We conduct ablation studies on the SA-Eval to assess the impact of the AT module and ASR-assisted prediction on performance, as shown in Table \ref{tab:ablation}.

First, on the \textit{easy} mode of SA-Eval, we observe similar performance across various models. %
This can be attributed to the non-overlap data in this mode, which leads to similar model behaviour when processing speech instructions and audio information.
The analysis in Section \ref{sec:ana_asr_ass} further supports this point. %

Second, our experimental results show that incorporating the AT Module enhances model performance compared to models without it on audio classification and audio question answering tasks.
Specifically, models with the AT Module, such as \tool and \tool without ASR-assisted prediction, outperform their counterparts lacking the module (i.e., \tool without AT Module and \tool without both) across most datasets.
For the audio captioning task, performance variations may arise from the fact that recognizing audio events does not directly influence captioning results.
Captioning performance is also influenced by factors such as the model's ability to summarize and generalize.

Finally, models utilizing ASR-assisted prediction show improved performance relative to those without it on most of the test sets, with particularly notable gains observed in the audio captioning task.
These findings highlight the benefits of integrating both the AT Module and ASR-assisted prediction into our model architecture.

\begin{table}
\small
\centering
\caption{Results for ASR-assisted prediction and instruction-following accuracy. ``IF'' stands for instruction following, ``Pred'' stands for prediction, and ``w/o'' refers to without.  \label{tab:ana_wer}}
\resizebox{0.5\textwidth}{!}{
\begin{tabular}{lccccccc}
    \toprule
    \multirow{2}{*}{\textbf{Model}} & \textbf{Mode of} & \textbf{ASR}  & \textbf{IF} \\
    & \textbf{SA-Eval} & WER $\downarrow$  & ACC $\uparrow$ \\
    \midrule
    \midrule
    \tool & \multirow{2}{*}{\textit{easy}} & 0.0  & 100.0 \\
    \quad w/o ASR-Assisted Pred  & & -  & 100.0 \\
    \midrule
    \tool & \multirow{2}{*}{\textit{hard}}  & 1.9  & 98.9 \\
    \quad w/o ASR-Assisted Pred  &  & -  & 98.3 \\
     \bottomrule
\end{tabular}
}
\end{table}

\subsection{Analysis of ASR-Assisted Prediction and Instruction-Following Ability}
\label{sec:ana_asr_ass}
To better understand the role of ASR-assisted prediction, we further analyze the ASR-assisted prediction results and the model's instruction-following ability.
We conduct analysis on the FSD50K test set using WER and accuracy metrics, as shown in Table \ref{tab:ana_wer}.

In the \textit{easy} mode, the ASR achieves a WER of zero and the instruction-following accuracy reaches 100\%, suggesting that the performance gap between \tool and Audio LLM$_\text{text}$ also stems from differences in audio comprehension, with speech instructions potentially influencing the model's understanding of audio inputs.

In contrast, in the \textit{hard} mode, \tool outperforms \tool w/o ASR-assisted prediction, showing higher instruction-following accuracy.
This highlights that ASR-assisted prediction enhances the model's ability to interpret speech instructions. %

\section{Related Work}
\paragraph{Speech/Audio LLMs}
Recent advancements in speech and audio LLMs have been significant. 
Numerous studies have investigated the development of speech/audio LLMs, which can be categorized into two types.

The first category of models \cite{chu2023qwen,tang2024salmonn,gong2024listen,deshmukh2023pengi,kong2024audio,chen-etal-2024-llast} typically uses an encoder to process input signals, where the encoder may be pre-trained on unpaired speech or audio data, or trained on downstream tasks such as ASR.
The encoder's output is then processed by an optional adaptor and fed into the LLM.
These models specify tasks through text instruction prompts and have demonstrated strong performance on various tasks.

The second category of models \cite{defossez2024moshi,fang2024llama,xie2024mini} focuses on seamless voice interaction.
These models usually utilize speech tokenization techniques to represent and process speech signals, so that speech and text data can be treated as similar types of data for training models.
However, these models rarely focus on understanding audio information.

In contrast to these approaches, \tool distinguishes itself by emphasizing both the understanding of audio content and the processing of speech instructions. It is specifically designed to generate appropriate responses by leveraging this dual understanding.

\paragraph{Benchmark Datasets for Speech/Audio LLMs}
The growing interest in speech/audio LLMs has led to the development of new benchmark datasets \cite{chen2024voicebench,wang2024audiobench,huang2024dynamic,huang2024dynamic2,ao2024sdeval,sakshi2024mmau,yang2024air,chen-etal-2024-beyond-single} aimed at evaluating these models from various perspectives. 
Many of these benchmarks \cite{wang2024audiobench,yang2024air,sakshi2024mmau,huang2024dynamic,huang2024dynamic2,chen-etal-2024-beyond-single} focus on testing models' performance only based on text instructions.

In contrast, VoiceBench \cite{chen2024voicebench} evaluates system performance through both synthetic and real spoken instructions. This benchmark assesses general knowledge, instruction-following ability, and safety compliance across diverse speaker and environmental conditions, with audio containing background sounds unrelated to the speech instruction itself. This design serves to test the model's robustness to noise.

Similarly, SD-Eval \cite{ao2024sdeval} measures spoken dialogue models across multiple dimensions, such as linguistic content, paralinguistic cues (emotion, accent, age), and environmental context, through four sub-tasks. The \textit{test-env} subset of SD-Eval, which contains 690 sentences, includes audio and speech instructions. Here, the model must use information from the audio to provide answers, though the focus is on daily conversations.

Unlike the above benchmarks, SA-Eval is specifically designed for audio-related tasks such as audio classification, captioning, and question answering.
It creates specific speech instructions based on these tasks and categorizes the difficulty into \textit{easy} mode and \textit{hard} mode according to real-life situations.

\section{Conclusions}
In this paper, we introduce \tool, a speech-oriented LLM that hears acoustic context. \tool combines the AT module and ASR-assisted prediction to effectively interpret audio cues and accurately understand spoken questions.
In addition, we present SA-Eval, a comprehensive benchmark designed to evaluate models on six test sets for audio classification, audio captioning, and audio question answering, across two levels of difficulty.

Our experimental results demonstrate that \tool performs on par with or outperforms baseline models, particularly in the \textit{hard} mode, highlighting its robustness across different scenarios.
Furthermore, the analysis reveals that the ASR-assisted prediction method enhances the model's ability to interpret speech instructions, while the AT module may play a crucial role in understanding the audio content of the input signal.

\section*{Limitations}
The limitations of \tool and SA-Eval are as follows:
First, \tool and SA-Eval focus on scenarios where the input is audio and the output is text, thereby restricting the output to text form.
Second, \tool and SA-Eval only address understanding tasks in the audio events, without considering music-related aspects.
Lastly, SA-Eval's evaluation currently focuses only on single-turn dialogues, limiting its ability to assess multi-turn complex dialogues.

\bibliography{ref}

\appendix

\section{Instruction Generation Details}
\label{apd:ins_gen}

First, we formulate five different base instructions for the audio captioning and audio classification tasks.
The content can be summarized by completing the specific task based on background sound. Subsequently, we utilize GPT-4o \cite{openai2024gpt4o} to paraphrase each instruction by 100 times, using the prompt: \textit{Generate 100 new versions of this sentence with synonyms: \{base instruction\}}.

As discussed in \ref{Section 3.3}, the generated instructions contain overly specific words that cannot be universally applied to all test cases.
For example, certain terms, such as \textit{harmonics} for background sound, are used in contexts where they do not apply.
Therefore, human annotators, who are the student volunteers, identify these words and delete the instructions that contain them.
Table \ref{tab:exam_inst} shows some examples of instructions for the audio classification and audio captioning tasks.

\begin{table*}
\centering
\caption{Examples of instructions for the audio classification and audio captioning tasks.\label{tab:exam_inst}}
\resizebox{\textwidth}{!}{
\begin{tabular}{ll}
    \toprule
    \textbf{Task}  & \textbf{Example} \\
    \midrule
    \midrule
    \multirow{5}{*}{\textbf{Audio Classification}} & Classify all the sounds in the background and let me know their labels. \\
    & Can you identify and provide the sound labels for all events from the background? \\
    & Please identify every audio event in the background and provide its label. \\
    & Could you tell me the names of the sound labels for all audio events in the background? \\
    & Could you identify and classify the audio events present in the background sound? \\
    \midrule
    \multirow{5}{*}{\textbf{Audio Captioning}} & Can you summarize the audio in the background with a caption? \\
    & Would you draft a caption describing the ambient sound and share it with me?\\
    & Could you create a caption that describes the background audio and let me know?\\
    & Write a description of the background audio.\\
    & Describe the background audio in a caption, and share it with me.\\
     \bottomrule
\end{tabular}
}
\end{table*}

\section{Mixing Algorithm of Audio and Speech Instruction}
Algorithm \ref{alg:1} shows the detailed steps of mixing audio and speech instruction.

\label{apd:mix_algo}
\begin{algorithm2e}[ht]
\footnotesize
\caption{Generating Mixture of Audio and Speech Instruction}\label{alg:1}
\KwData{ \\
    $s_s$: Input signals of the speech instruction \\
    $s_a$: Input signals of the audio \\
    $l_a$: Length of the audio  \\
    $l_s$: Length of the speech instruction \\
    $m$: \textit{easy} or \textit{hard} \\
}

\uIf{m is hard}{
    \tcp{If overlap, selecting start time and padding the shorter one}
    \eIf{$l_a < l_s$}{
        $t_{s} \gets randint(0, l_s - l_a)$\;
        $s_a \gets pad(s_a, (t_{s}, l_s - l_a - t_{s}))$\;
    }{
        $t_{s} \gets randint(0, l_a - l_s)$\;
        $s_s \gets pad(s_s, (t_{s}, l_a - l_s - t_{s}))$\;
    }
}\uElseIf{m is easy}{
    \tcp{If not overlap, padding both audio and speech}
    \eIf{$\text{random()} > 0.5$}{
        $s_a \gets pad(s_a, (l_s, 0))$\;
        $s_s \gets pad(s_a, (0, l_a))$\;
    }{
        $s_a \gets pad(s_a, (0, l_s))$\;
        $s_s \gets pad(s_a, (l_a, 0))$\;
    }
}
\tcp{Selecting loudness for speech instruction and audio}
\uIf{m is hard}{
    $v_a \gets randint(-23, -20)$\;
    $v_s \gets randint(-33, -30)$\;
}\uElseIf{m is easy}{
    $v_a \gets randint(-38, -20)$\;
    $v_s \gets randint(-33, -25)$\;
}

\tcp{Setting loudness and clipping if needed}
$s_a \gets set\_loudness(s_a, v_a)$\;
$s_s \gets set\_loudness(s_s, v_s)$\;

\tcp{Mixing and getting mixture speech $s_m$}
$s_m \gets s_s + s_a$\;
\textbf{Output:} speech mixture $s_m$
\end{algorithm2e}

\section{Training Set Statistics}
Table 6 presents the statistics of the training data.
Except for MusicQA \cite{liu2024music} and ClothoAQA \cite{lipping2022clotho}, which use the original question as the instruction, the rest are sourced from the OpenAQA dataset \cite{gong2024listen}.
\label{apd:trainin_set}

\begin{table*}
\small
\centering
\caption{Statistics of the training set.\label{tab:train_dataset}}
\begin{tabular}{lcccccc}
    \toprule
    \textbf{Dataset}  & \textbf{\# QA Pairs}  & \textbf{\# Audio Samples} &  \textbf{Duration (hrs)} \\
    \midrule
    \midrule
    FSD50K \cite{fonseca2021fsd50k} & 441,119 & 40,128 & 79.8  \\
    AudioSet-2M \cite{gem2017audioset} & 394,865 & 394,865 & 1087.3 \\
    AudioSet-20K \cite{gem2017audioset} & 194,970 & 18,301 & 50.4 \\
    AudioSet Strong \cite{9414579} & 1,276,793 & 99,865 & 273.7 \\
    AudioCaps \cite{kim2019audiocaps} & 491,326 & 45,639 & 125.0 \\
    FreeSound \cite{10.1145/2502081.2502245} & 768,822 & 86,845 & 107.8 \\
    Sound Bible \cite{soundbible} & 16,225 & 1,104 & 2.2 \\
    VGGSound \cite{chen2020vggsound} & 1,104,020 & 181,700 & 503.7 \\
    Clotho V2 \cite{9052990} & 117,406 & 4,780 & 29.9 \\
    MusicQA \cite{liu2024music} & 104,106 & 12,542 & 76.2 \\
    ClothoAQA \cite{lipping2022clotho} & 5,666 & 1,174 & 7.4 \\
    \midrule
    Overall & 4,915,318 & 770,351 & 2,023.6 \\
     \bottomrule
\end{tabular}
\end{table*}

\section{Evaluation Judgment}
\label{apd:eval_jud}

\begin{figure*}
    \centering
    \includegraphics[width=0.8\textwidth]{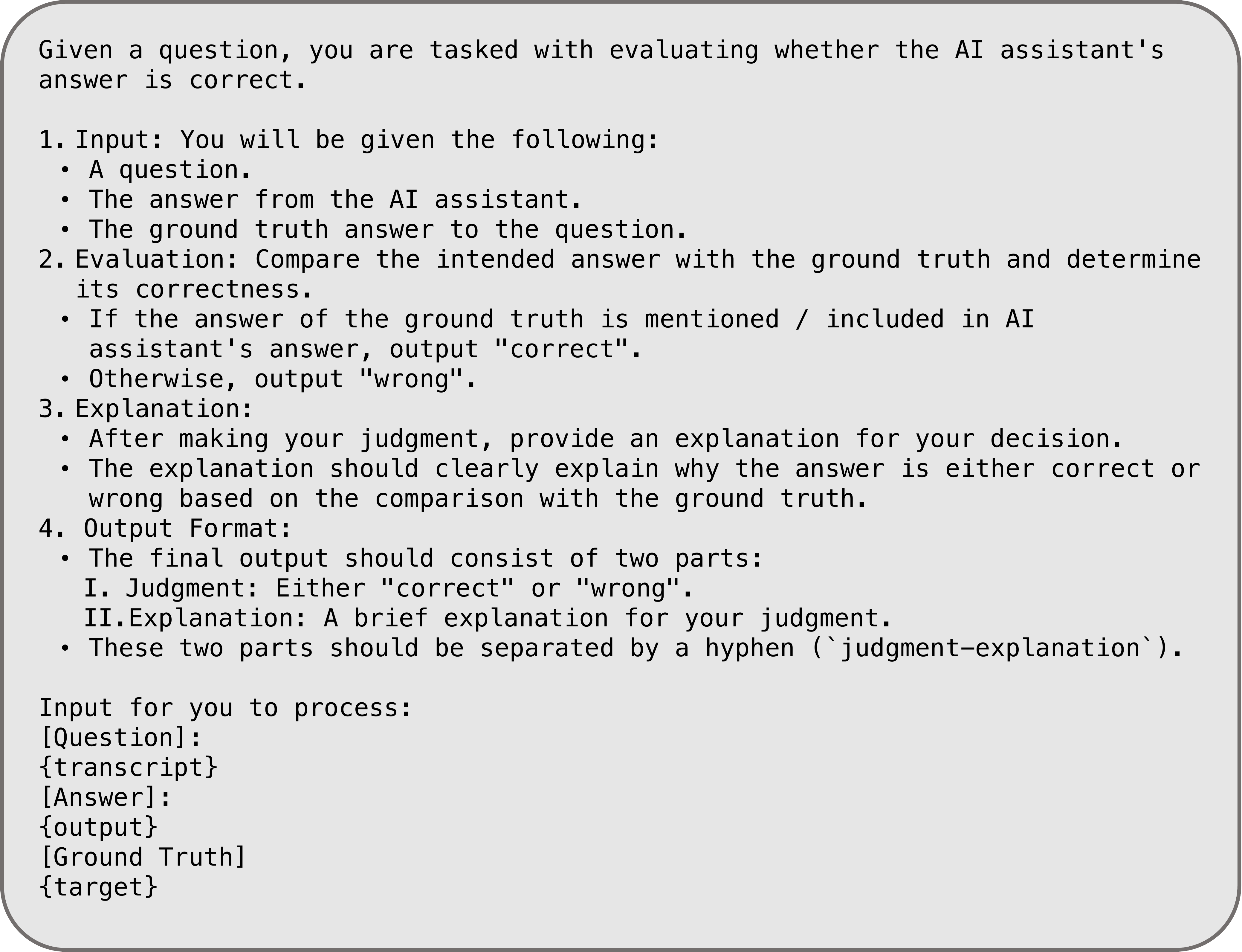}
    \caption{Evaluation judgment prompt of VGGSound. \label{fig:vgg_prompt}}
\end{figure*}

\begin{figure*}
    \centering
    \includegraphics[width=0.8\textwidth]{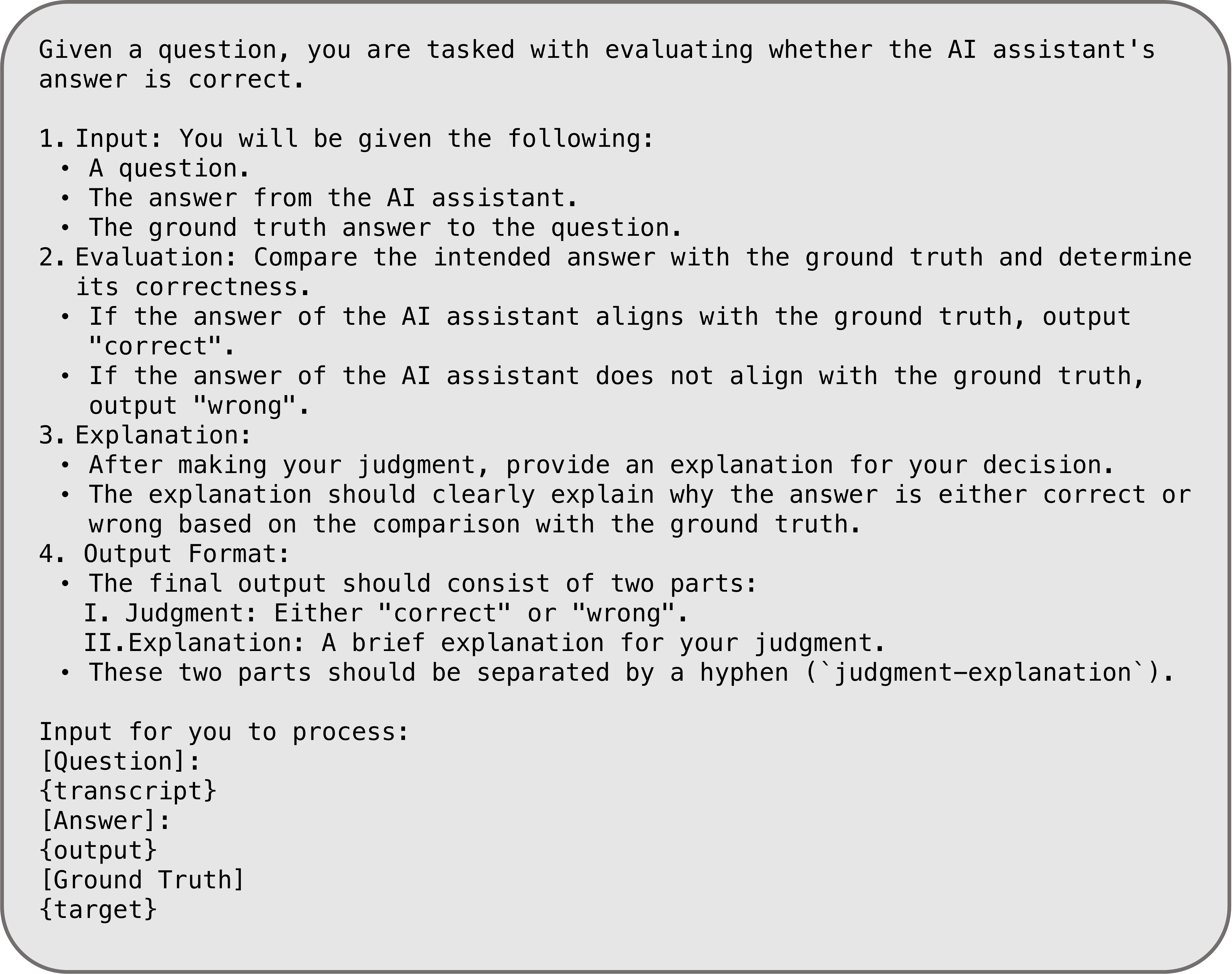}
    \caption{Evaluation judgment prompt of Clotho-AQA. \label{fig:aqa_prompt}}
\end{figure*}

For single-label tasks, VGGSound and Clotho-AQA, an English normalizer is first applied to both the ground-truth and prediction outputs. The outputs are considered a correct answer if they are found to be identical by character. Otherwise, the o1-mini is utilized to evaluate the prediction's correctness, with prompts shown in Figure \ref{fig:vgg_prompt} and Figure \ref{fig:aqa_prompt}.

\end{document}